\documentclass[10pt]{iopart}

\usepackage{graphicx}
\usepackage[export]{adjustbox}
\graphicspath{{./figs/}}

\usepackage{bm}
\usepackage{iopams}

\newcommand{\pdv}[2]{\frac{\partial {#1}}{\partial {#2}}}
\newcommand{\pdvi}[2]{\partial {#1}/\partial {#2}}
\newcommand{\va}[1]{\mathbf{#1}}
\newcommand{\abs}[1]{\left|{#1}\right|}
\newcommand{\bmcal}[1]{\boldsymbol{\mathcal{#1}}}
\newcommand{\ev}[3]{\langle{#1}|{#2}|{#3}\rangle}
\newcommand{\dd}[1]{\mathrm{d}{#1}}
\newcommand{\dds}[1]{\mathrm{d}^2{\va{#1}}}
\newcommand{\st}[1]{|{#1}\rangle}

\newcommand{\ie}{i.e.~}
\newcommand{\cf}{c.f.~}
\newcommand{\eg}{e.g.~}

\usepackage[normalem]{ulem}
\usepackage[usenames,dvipsnames]{xcolor}

\begin{document}
\title[Nonlinear optical selection rules]{Nonlinear optical selection rules of excitons in monolayer transition metal dichalcogenides}

\author{A Taghizadeh$^{1}$ and T G Pedersen$^{1,2}$}

\address{$^1$ Department of Physics and Nanotechnology, Aalborg University, DK-9220 Aalborg {\O}st, Denmark}
\address{$^2$ Center for Nanostructured Graphene (CNG), DK-9220 Aalborg {\O}st, Denmark}

\ead{ata@nano.aau.dk}

\begin{abstract}
We propose an analytical approach for calculating the linear and nonlinear optical (NLO) responses of monolayer transition metal dichalcogenides (TMDs) including excitonic effects. An effective Hamiltonian reproducing the trigonal warping (TW) of the energy dispersion, is used to derive analytical expressions for excitonic matrix elements. Based on this approach, we provide an informative diagram, which encompasses all excitonic selection rules. The diagram enables us to identify main transitions for the first-, second- and third-order optical responses. As a case study, we calculate the optical conductivity and second/third harmonic generation responses of monolayer MoS$_2$ and demonstrate that the analytical approach accurately reproduces the spectra obtained using the Bethe-Salpeter equation (BSE). Therefore, the analytical approach provides an efficient and reliable method for calculating the NLO spectra as accurate as the full BSE method. Moreover, it enables us to obtain valuable physical insight into the fundamental transitions responsible for individual resonances, which is not straightforward in the full BSE method. 
\end{abstract}

\submitto{\TDM}
\maketitle
\ioptwocol

\section{Introduction}
The interest in the NLO response of 2D materials has recently grown drastically due to their large nonlinear coefficients, which facilitate a vast range of applications in optoelectronic devices \cite{Sun2016, Autere2018} as well as fundamental physics, \eg nonlinear charge and spin Hall effects \cite{Zhang2018b, Eginligil2015, Taghizadeh2019}. 
Various NLO processes in semi-metallic graphene have been extensively studied both theoretically and experimentally \cite{Hendry2010, Hong2013, Rostami2017, Hipolito2017, Baudisch2018, Jiang2018}. In centrosymmetric graphene, even-order NLO processes vanish in the dipole approximation \cite{Hong2013}. In contrast, the intrinsic broken inversion symmetry in monolayer TMDs leads to non-vanishing even-order NLO responses. Furthermore, strong excitonic effects due to reduced screening as well as huge spin-orbit coupling (SOC) in monolayer TMDs make their optical and electronic properties unique \cite{Schaibley2016}.

The linear optical response of monolayer TMDs has been extensively studied theoretically using the BSE in combination with single-particle band structures obtained by ab-initio techniques \cite{Qiu2013, Ugeda2014, Thygesen2017} or tight-binding (TB) methods \cite{Berghauser2014, Wu2015, Ridolfi2018}. Beside these rigorous calculations, analytical approaches have also been developed by employing the so-called massive Dirac Hamiltonian to obtain physical insight \cite{Berghauser2014, Wu2015, Xiao2012, Chaves2017}. Employing the analytical approach not only reduces the computational burden of excitonic optical response calculations but also leads to the identification of excitonic selection rules for the linear optical response in monolayer TMDs \cite{Cao2018, Zhang2018}. Optical selection rules are fundamental principles, which determine allowed and forbidden transitions mediated by light. These rules are imposed by various symmetries of the system such as time-reversal, spatial or rotational symmetries. For instance, the brightness of $s$ and darkness of $p$ excitons in the linear optical spectrum of monolayer TMDs have been explained by selection rules \cite{Cao2018, Zhang2018}. Note that although excitons in monolayer TMDs are of Wannier-Mott type, which are also found in conventional semiconductors, they possess different selection rules due to the chirality of the band structure at the bandgap \cite{Zhang2018}.
Regarding the NLO response, a few calculations for monolayer TMDs can be found, in which the BSE solution is employed for computing the second harmonic susceptibilities \cite{Trolle2015, Gruning2014, Wang2015}. While the linear optical response originates only from the transition between the ground state and excited states, the NLO responses depend also on the transition between excited states, \ie inter-excitonic transitions. Therefore, more complicated selection rules are expected to govern NLO processes. The NLO selection rules have been studied qualitatively in Ref.~\cite{Xiao2015} and quantitatively using a Wannier model (with $1/r$ Coulomb potential) in a recent work \cite{Soh2018}. Nonetheless, the qualitative work can not provide any estimation for the excitonic matrix elements, while the Wannier model fails to accurately capture the nonlinear excitonic spectra due to the missing Bloch band curvature \cite{Zhang2018}. Therefore, a systematic study of excitonic selection rules and their effects on the NLO response including the Bloch band signature monolayer TMDs is missing.

In this work, we develop an analytical approach for systematic study of the NLO response in monolayer TMDs. By using an effective Hamiltonian, which includes the TW term, allowed and forbidden transitions between excitonic states are identified in the dipole approximation. A diagrammatic method is then suggested, as a means for predicting the dominant transitions and their origin for NLO processes up to any required order. We apply the proposed analytical approach to monolayer MoS$_2$ as a case study and confirm the validity and accuracy of the analytical approach by comparing with the full BSE results. 
This paper is organized as follows. We begin by reviewing the theoretical framework for the NLO response in monolayer TMDs using the effective TW Hamiltonian. Based on the developed framework, optical selection rules for the first-, second- and third-order processes are discussed in \sref{sec:Selection}. In \sref{sec:Numerics}, the linear and NLO conductivities computed for monolayer MoS$_2$ are reported. Finally, a summary of the main findings is provided in \sref{sec:Conclusion}. A set of appendices explain the details of the derived expressions.

\section{Theoretical framework \label{sec:Theory}}
In this section, we present the theoretical framework for calculating the NLO response of monolayer TMDs using our analytical approach. First, general expressions for the optical conductivity (OC) and second/third harmonic generation (SHG/THG) are reviewed. Then, the TW Hamiltonian for monolayer TMDs is introduced. Finally, analytical expressions for the required excitonic matrix elements are presented. It should noted that throughout the text, all vectors and tensors are indicated by bold letters, and the single-particle/many-body operators and matrix elements are denoted by lowercase/uppercase letters, respectively.

\subsection{Excitonic nonlinear optical response}
The linear and NLO responses of a material can be characterized by linear and NLO conductivities (or equivalently susceptibilities) \cite{Boyd2008}. In the present work, we assume that the material is excited by a normal-incident monochromatic electric field written as $\bmcal{E}(t)=\bmcal{E}_{\omega}\exp(-i\omega t)+\mathrm{c.c.}$, where $\bmcal{E}_{\omega}=\mathcal{E}_{\omega}^x \bm{e}_x+\mathcal{E}_{\omega}^y \bm{e}_y$ is the electric field phasor ($\bm{e}_x$ and $\bm{e}_y$ are unit vectors along the $x$- and $y$-directions, respectively). 
The light-matter interaction is then studied using a density matrix approach, which has been explained in detail in our previous works, \cf Refs. \cite{Pedersen2015, Taghizadeh2018}. In this approach, a set of coupled equations for the reduced density matrix $\rho_{ij\va{k}}=\ev{0}{\hat{c}_{i\va{k}}^\dagger\hat{c}_{j\va{k}}}{0}$ is derived, where $\st{0}$ denotes the many-body ground state. Furthermore, $i$ and $j$ are band indices, $\va{k}$ is the wavevector, and $\hat{c}^\dagger$ ($\hat{c}$) is the fermionic creation (annihilation) operator. Neglecting exchange terms, the dynamical equations for $\rho_{ij\va{k}}$ for a two-band semiconductor within the mean-field approximation are presented in Eqs.~(\ref{eq:DensityDiffEqA}) and (\ref{eq:DensityDiffEqB}) of \ref{sec:AppendixA}. These equations are then solved perturbatively up to any required order of the field at vanishing temperature, and the $N^{th}$-order density matrix elements $\rho_{ij\va{k}}^{(N)}$ are determined. In the present work, we only focus on the NLO processes of second/third harmonic generation for simplicity, and the required density matrices are found in Eqs.~(\ref{eq:DensityFinalA})-(\ref{eq:DensityFinalF}).

Upon determining the density matrix, the $N^{th}$ harmonic current density is evaluated as $\va{J}^{(N)}(t)=\mathrm{Tr}[\hat{\rho}^{(N)}\hat{\va{J}}]$, where the charge current density operator reads $\hat{\va{J}}=-e\hat{\va{\Pi}}/(mA)$ (for a crystal area $A$) \cite{Taghizadeh2018}. Here, $\hat{\va{\Pi}}$ is the Heisenberg momentum defined as $\hbar\hat{\va{\Pi}}=im[\hat{\mathcal{H}}_0,\hat{\va{R}}]$, where $\hat{\mathcal{H}}_0$ and $\hat{\va{R}}$ are the total unperturbed Hamiltonian and many-body position operators, respectively \cite{Taghizadeh2018}. In turn, the $N^{th}$ harmonic generated current density reads
\begin{equation}
	\va{J}^{(N)}(t) = \boldsymbol{\sigma}^{(N)} \underbrace{\bmcal{E}_{\omega}...\bmcal{E}_{\omega}}_{N\,\,\mathrm{times}} \exp(-iN\omega t) + \mathrm{c.c.} \, ,
\end{equation}
where $\boldsymbol{\sigma}^{(N)}$ is the $N^{th}$-order conductivity tensor ($N \geq 1$). Note that linear and NLO susceptibilities, $\boldsymbol{\chi}^{(N)}$
are related to their corresponding conductivities by $\boldsymbol{\sigma}^{(N)}=-iN\omega\epsilon_0\boldsymbol{\chi}^{(N)}$. The expressions of the first-, second-, and dominant part of the third-order conductivity tensors at vanishing temperature are given by \cite{Taghizadeh2018} 
\numparts \begin{eqnarray}
	\label{eq:Conductivity1st}
	&\fl \sigma_{\alpha\beta}^{(1)} = C_1 \sum_m \bigg[ \frac{\Pi_{0m}^\alpha X_{m0}^\beta}{\hbar\omega-E_m}-\frac{\Pi_{m0}^\alpha X_{0m}^\beta}{\hbar\omega+E_m} \bigg] \, , \\
	\label{eq:Conductivity2nd}
	&\sigma_{\eta\alpha\beta}^{(2)} = C_2 \sum_{n,m} \bigg[ \frac{\Pi_{0n}^\eta Q_{nm}^\alpha X_{m0}^\beta}{(2\hbar\omega-E_n)(\hbar\omega-E_m)} + \nonumber \\
	&\frac{\Pi_{n0}^\eta Q_{mn}^\alpha X_{0m}^\beta}{(2\hbar\omega+E_n)(\hbar\omega+E_m)} - \frac{X_{0n}^\alpha \Pi_{nm}^\eta X_{m0}^\beta}{(\hbar\omega+E_n)(\hbar\omega-E_m)} \bigg] \, , \\
	\label{eq:Conductivity3rd}
	&\sigma_{\mu\eta\alpha\beta}^{(3)} = C_3 \sum_{l,n,m} \bigg[ \frac{\Pi_{0l}^\mu Q_{ln}^\eta Q_{nm}^\alpha X_{m0}^\beta}{(3\hbar\omega-E_l)(2\hbar\omega-E_n)(\hbar\omega-E_m)} \nonumber \\ 
	& \qquad \qquad -\frac{\Pi_{l0}^\mu Q_{nl}^\eta Q_{mn}^\alpha X_{0m}^\beta}{(3\hbar\omega+E_l)(2\hbar\omega+E_n)(\hbar\omega+E_m)} + \nonumber \\ 
	&\frac{X_{0l}^\alpha X_{m0}^\beta}{(\hbar\omega+E_l)(\hbar\omega-E_m)} \Big( \frac{Q_{ln}^\eta \Pi_{nm}^\mu}{2\hbar\omega+E_n} - \frac{\Pi_{ln}^\mu Q_{nm}^\eta}{2\hbar\omega-E_n} \Big) \bigg] \, . 
\end{eqnarray} \endnumparts 
Here, $C_3 \equiv eC_2 \equiv e^2C_1 \equiv -e^4 \hbar/(m^2 A)$, $\mu,\eta,\alpha,\beta=\{x,y\}$ are the Cartesian coordinates, and $E_n$ is the exciton energy. In addition, the excitonic matrix elements are defined as $\Pi_{0n}^\alpha = (\Pi_{n0}^\alpha)^* \equiv -iE_n X_{0n}^\alpha$, $\Pi_{nm}^\alpha \equiv i(E_n-E_m)Q_{nm}^\alpha$, and 
\numparts \begin{eqnarray}
	\label{eq:ExcitonMatrix}
	&X_{0n}^\alpha = (X_{n0}^\alpha)^* \equiv -\sum_{\va{k}} \psi_{\va{k}}^{(n)} \frac{p_{vc\va{k}}^\alpha}{i\varepsilon_{cv\va{k}}} \, , \\ 
	&Q_{nm}^\alpha \equiv i \sum_{\va{k}} \psi_{\va{k}}^{(n)*} \big[\psi_{\va{k}}^{(m)} \big]_{;k_\alpha} \, , 
\end{eqnarray} \endnumparts
where $\varepsilon_{cv\va{k}} \equiv \varepsilon_{c\va{k}}-\varepsilon_{v\va{k}}$ (indices $c$ and $v$ imply conduction and valence bands, respectively), and $p_{vc\va{k}}^{\alpha} \equiv \ev{v\va{k}}{\hat{p}_\alpha}{c\va{k}}$ are the transition energy and momentum matrix element between single-particle states, respectively. Moreover, $\psi_{\va{k}}^{(n)}$ is the exciton wavefunction projected onto a band-to-band transition, \ie the exciton state is expressed as $\st{\psi^{(n)}} = \sum_{\va{k}} \psi_{\va{k}}^{(n)} \hat{c}_{c\va{k}}^\dagger \hat{c}_{v\va{k}} \st{0}$ \cite{Benedict1999} (the exciton center-of-mass motion is neglected due to the negligible photon momentum). The summations over $\va{k}$ should be understood as integrals over the Brillouin zone (BZ), \ie $(2\pi)^2 \sum_\va{k} \rightarrow A \int_{\mathrm{BZ}} \dds{\va{k}}$. Finally, $[\psi_{\va{k}}^{(n)}]_{;k_\alpha} \equiv \partial \psi_{\va{k}}^{(n)}/\partial k_\alpha-i(\Omega_{cc\va{k}}^{\alpha}-\Omega_{vv\va{k}}^{\alpha}) \psi_{\va{k}}^{(n)}$ denotes the generalized derivative of the exciton wavefunction with the Berry connections $\Omega_{nm\va{k}}^{\alpha} \equiv \ev{n\va{k}}{i\partial/\partial k_\alpha}{m\va{k}}$. The generalized derivative stems from the intraband part of the position operator \cite{Aversa1995, Taghizadeh2017}. Equations~(\ref{eq:Conductivity1st}) and (\ref{eq:Conductivity2nd}) have been reported in our previous work \cite{Taghizadeh2018} while Eq.~(\ref{eq:Conductivity3rd}) can be obtained in a similar manner by extending the general approach to the third order in the electric field. Although these expressions are obtained for a two-band semiconductor, they can be used for multi-band semiconductors by redefining the $Q_{nm}^\alpha$ matrix elements (see Ref.~\cite{Taghizadeh2018} for more details). Note that, with a minor modification, these expressions can also be employed for calculating the nonlinear spin conductivities \cite{Taghizadeh2019}.

To determine excitonic energies and wavefunctions, we solve the BSE \cite{Benedict1999, Albrecht1998}, given by  
\begin{equation}
\label{eq:FullBSE}
	\varepsilon_{cv\va{k}} \psi_{\va{k}}^{(n)} - \sum_{\va{k}'} V_{cv}^{\va{k},\va{k}'} \psi_{\va{k}'}^{(n)} = E_n\psi_{\va{k}}^{(n)} \, ,
\end{equation}
where $V_{cv}^{\va{k},\va{k}'}$ is the direct Coulomb matrix elements defined in Eq.~(\ref{eq:CoulombMatrix}) of \ref{sec:AppendixA}. The Coulomb matrix elements include band signatures through the Bloch overlaps, $\langle n\va{k}|n\va{k}'\rangle$. For 2D materials, a screened Coulomb interaction in the Keldysh form is employed, which accurately describes the excitonic spectrum \cite{Wu2015, Taghizadeh2018, Cudazzo2011, Trolle2017}. The BSE solutions contain both the bound and unbounded excitons. 

\subsection{Trigonal warping Hamiltonian for monolayer TMDs}
\begin{figure}[t]
	\centering
	\includegraphics[width=0.49\textwidth]{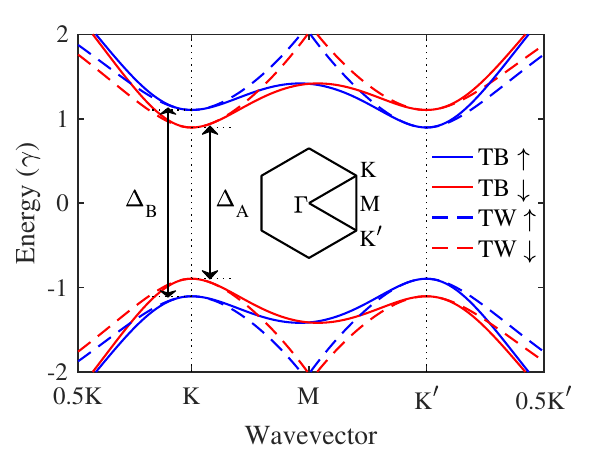}
	\caption[Band structure]{Typical band structure of monolayer TMD near the bandgap obtained using the full TB (solid lines) or TW Hamiltonian (dashed-lines) for $\gamma=\Delta=50\lambda$. Blue and red lines show spin-up ($s=1$) and -down ($s=-1$) bands, respectively. The inset shows the BZ and its high symmetry points. }
	\label{fig:BandStructure}
\end{figure}

It is well-known that the valence and conduction bands of monolayer TMDs in the vicinity of the Dirac points are mainly formed by $(|d_{x^2-y^2}\rangle+|d_{xy}\rangle)/\sqrt{2}$ and $|d_{z^2}\rangle$ orbitals of the metal atom, respectively \cite{Xiao2012, Kormanyos2013, Konabe2014}. Hence, we can construct an effective 2-by-2 tight-binding (TB) Hamiltonian \cite{Taghizadeh2019, Berghauser2014}, which reproduces the low-energy band structure of monolayer TMDs. Including the SOC, the TB Hamiltonian in $\va{k}$-space reads \cite{Taghizadeh2019, Qiao2012, Pedersen2018}
\begin{equation}
	\label{eq:TBHamiltonian}
	H_0(\va{k}) = \left[ \begin{array}{cc} \Delta+\lambda g(\va{k})s & -\gamma f^*(\va{k}) \\ -\gamma f(\va{k}) & -\Delta-\lambda g(\va{k})s \end{array} \right] \, ,
\end{equation}
where $s=\pm1$, $\Delta$, $\gamma$ and $\lambda$ are the spin index, on-site energy, effective hopping and SOC strength, respectively. The wavevector-dependent functions read
\numparts \begin{eqnarray}
&f(\va{k}) \equiv e^{ik_x a/\sqrt{3}} + 2 e^{-ik_x a/(2\sqrt{3})} \cos(k_y a/2) \, , \\
&g(\va{k}) \equiv 2\sin(k_y a)-4\sin(k_y a/2) \cos(k_x a/\sqrt{3}) \, ,
\end{eqnarray} \endnumparts
with the lattice constant $a$. A typical band structure obtained for this Hamiltonian is illustrated in Fig.~\ref{fig:BandStructure}. The SOC lifts the degeneracy of the bands at each valley, while the time-reversal symmetry (TRS) relates the bands in separate valleys, \ie spin-up states at the K valley are degenerate with spin-down states at the K$'$ valley. 


The massive Dirac Hamiltonian can be derived from Eq.~(\ref{eq:TBHamiltonian}) by a Taylor expansion of $f(\va{k})$ and $g(\va{k})$ around the Dirac points to linear order in $\va{k}$. Although this approximate Hamiltonian has proven useful for characterizing the linear optical properties of TMD monolayers, it fails to account for any even-order nonlinear response (in the dipole approximation) due to the presence of full rotation symmetry, $C_\infty$ \cite{Taghizadeh2019}. This issue can be resolved by including terms up to second order in $\va{k}$ for $f(\va{k})$, which leads to restored threefold rotation symmetry, $C_3$, and the TW of isoenergy contours \cite{Kormnyos2015}. Hence, we form the TW Hamiltonian by expanding $g(\va{k})$ and $f(\va{k})$ at the K/K$'$ valleys as $g(\va{k}) \approx 3\sqrt{3}\tau$ and 
\begin{equation}
\label{eq:TWFunction}
	f(\va{k}) \approx \frac{\sqrt{3}}{2} e^{i\pi/3} \big[-i(\kappa_x-i\kappa_y\tau) + \zeta(\kappa_x+i\kappa_y\tau)^2 \big] \, .
\end{equation}
Here, $\kappa_\alpha\equiv a(k_\alpha-K_\alpha)$ is the dimensionless wavevector measured with respect to K or K$'$, $\zeta \equiv \sqrt{3}/12$ multiplies the trigonal warping term, and $\tau=\pm 1$ is the valley index. Note that the second-order term in $\kappa$ in the expansion of $g(\va{k})$ is neglected here, since its contribution is much smaller than the equivalent term in $f(\va{k})$ due to $\lambda \ll \gamma$. The band structure obtained from the TW Hamiltonian is compared with the TB band structure in Fig.~\ref{fig:BandStructure}. Although the TW band structure deviates from the TB band structure away from K or K$'$, it accurately reproduces the optical response at low photon frequency \cite{Taghizadeh2019}. Note that the band structure in Fig.~\ref{fig:BandStructure} is slightly shifted compared to the typical band structure of monolayer TMDs, where the SOC mainly lifts the valence band degeneracy. However, this vertical shift of the band structure is not important in our calculations, since only energy differences (transitions) matter.
Using the TW Hamiltonian, we derive analytical expressions for the momentum matrix elements and Berry connections to first order in $\zeta$ as presented in Eqs.~(\ref{eq:TransitionEnergy})-(\ref{eq:yBerry}) of \ref{sec:AppendixB}. These equations have previously been employed to obtain analytical expressions for the linear and NLO responses in the independent-particle approximation, which are in excellent agreement with the full TB results \cite{Taghizadeh2019}.

\subsection{Analytical approach for excitonic optical response}
The impact of excitons on the optical response can be accurately studied by incorporating the BSE solution. Although solving the BSE for a given single-particle band structure is possible, in principle, it is a numerically demanding practice. Furthermore, the physical interpretation of the optical response is not straightforward. Fortunately, the approximate solution of the BSE using the massive Dirac Hamiltonian provides not only accurate estimates of the excitonic energies and wavefunctions but also valuable physical insight \cite{Wu2015, Zhou2015}.
In the massive Dirac approximation, the BSE is transformed to a modified Schr\"{o}dinger equation for a 2D Hydrogen-like atom \cite{Zhou2015} hereafter referred to as the \textit{analytical BSE}, see Eq.~(\ref{eq:AnalyticBSE}) in \ref{sec:AppendixC}. Due to the  full rotational symmetry, the solution of the analytical BSE yields an exciton wavefunction with a distinct angular momentum, given by $\psi_{\va{k}}^{(n)} = \exp(il\theta ) \phi_{l}^{(n)}(\kappa)$. Here, $\theta$ and $\kappa$ are the phase and magnitude of dimensionless wavevector, respectively, $l$ denotes the angular momentum of the exciton, and $\phi_{l}^{(n)}(\kappa)$ is the radial part of exciton wavefunction. Note that the angular momentum $l$ is defined in reciprocal space rather than real coordinate space.


In the BSE, the band overlaps $\langle n\va{k}|n\va{k}' \rangle$, appearing in the integrand, are sensitive to the chosen phase of Bloch states. Although the choice of phase does not influence any measurable quantity such as the optical response, it can change the interpretation of angular momenta in the solutions of the analytical BSE. In analogy to the Hydrogen atom orbitals, Bloch phases can be chosen in such a way that the excitons with $l=-2,-1,0,+1,+2$ correspond to orbitals $d^-,p^-,s,p^+,d^+$, respectively \cite{Cao2018, Zhang2018, Srivastava2015}. In the present work, however, we have chosen a different phase convention for the Bloch states, which leads to a $-1$ shift of the exciton angular momenta at the K valley, \ie the $l=-3,-2,-1,0,+1$ states are labeled as $d^-,p^-,s,p^+,d^+$ orbitals, respectively. At the K$'$ valley, the angular momentum shift is $+1$, while the plus and minus orbitals interchange their roles due to the TRS, \eg $p^-$ ($l=-2$) at the K valley is degenerate with $p^+$ ($l=+2$) at the K$'$ valley. Transforming from our phase convention to the conventional Hydrogen atom phase convention is straightforward by relabeling excitons, yet it simplifies our analytical expressions considerably. Note that, analogous to the Hydrogen orbitals, excitons are labeled as $ns$, $np^\pm$, $nd^\pm$, etc, where $n$ is the principal quantum number starting from 1, 2 and 3 for $s$, $p^\pm$ and $d^\pm$ orbitals, respectively.

Besides the issue of exciton labeling, the calculated excitonic spectrum of the analytical BSE deviates from the conventional Rydberg series known for Hydrogen atoms due to the presence of Bloch band signatures and screened Coulomb interaction. In particular, the degeneracy between orbitals with the same principal quantum number is lifted. This has several implications for the energetic ordering of exciton states. 
Firstly, for the same principal quantum number, orbitals with lower angular momenta have higher energies (or smaller binding energies), \eg $E_{3d^\pm} < E_{3p^\pm} < E_{3s}$ \cite{Ye2014}. This can be explained by considering the screening property of the Keldysh potential. In fact, the 2D Coulomb interaction is only effectively screened for charges with small in-plane separation \cite{Cudazzo2011, Trolle2017}.  The Keldysh potential has a weak logarithmic divergence at short distances, while it approaches the $1/r$ dependence of the conventional Coulomb potential at large distances \cite{Cudazzo2011}. For excitons with lower angular momenta, the wavefunction is confined closer to the origin, \ie the hole and electron are closer to each other. Therefore, excitons with lower angular momenta experience less Coulomb interaction due to the pronounced short-range screening of the Keldysh potential and, hence, possess smaller binding energies.
Secondly, orbitals with positive and negative angular momenta, \eg $3p^+$ and $3p^-$, are not degenerate in a single valley, due to the non-zero Berry curvature \cite{Srivastava2015}. Nonetheless, due to the TRS, orbitals with positive angular momenta at the K valley are degenerate with the negative ones at the K$'$ valley and vice versa. Note that, if the TW Hamiltonian is used for the analytical BSE, a group of orbitals with angular momenta $l,l\pm3,l\pm6,\cdots$ are coupled \cite{Zhang2018}, \eg $s$ states are coupled to $f^\pm$ states. This coupling is typically weak ($\zeta<1$) and, hence, ignored here, which enables us to derive analytical expressions for the excitonic matrix elements.


\begin{figure*}[t]
	\centering
	\includegraphics[width=0.8\textwidth]{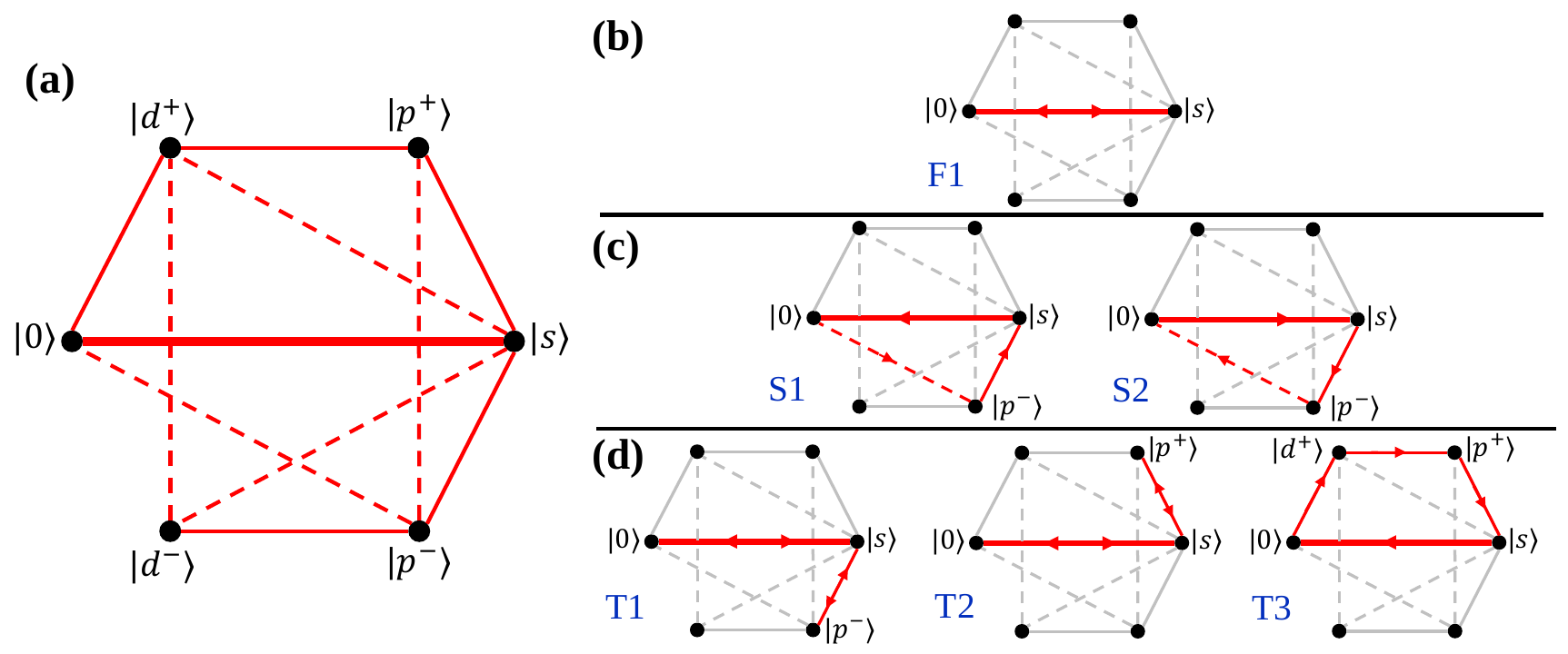}
	\caption[Schematic]{Schematic diagram of optical selection rules at the K valley. Black dots represent the ground state, $\st{0}$, and excited states, $\st{s}$, $\st{p^\pm}$ and $\st{d^\pm}$. Solid/dashed lines represent the possible transitions between the states, which are allowed without/with the TW. The thicker line for coupling between $\st{0}$ and $\st{s}$ states indicates larger matrix elements compared to the rest. Note that the transition may be in both directions, \eg $\st{0} \rightarrow \st{s}$ or $\st{s} \rightarrow \st{0}$. The dominant transition paths for the (b) first-, (c) second-, (d) third-order processes are shown in red. Note that a similar diagram can readily be provided for the K$'$ valley by interchanging the plus and minus states.  }
	\label{fig:Schematic}
\end{figure*}

Combining the analytical BSE solution with the matrix elements obtained using the TW Hamiltonian, \ie Eqs.~(\ref{eq:TransitionEnergy})-(\ref{eq:yBerry}), both linear and NLO responses can be determined. This is done by performing the angular integral ($\theta$) for the exciton matrix elements between the ground state $\st{0}$ and the excited states $\st{n}$, \ie $X_{0n}^\alpha$, or between the different excited states, \ie $Q_{nm}^\alpha$. Allowed and forbidden transitions are identified based on the properties of these matrix elements.  
The analytical expressions for $X_{0n}^x$ at the K valley for $s$, $p^\pm$ and $d^\pm$ excitons are given by
\numparts \begin{eqnarray}
	\label{eq:SelectionXs}
	&X_{0n}^x(s) = X_0 \int_0^\infty \phi_{-1}^{(n)} \frac{\eta_\kappa+1}{i\varepsilon_\kappa} \kappa \dd{\kappa} \, , \\
	\label{eq:SelectionXpp}
	&X_{0n}^x(p^+) = 0 \, , \\
	\label{eq:SelectionXpn}
	&X_{0n}^x(p^-) =  X_0\zeta \int_0^\infty \phi_{-2}^{(n)} \frac{\eta_\kappa^3+4\eta_\kappa+3}{2\varepsilon_\kappa} \kappa^2 \dd{\kappa} \, , \\
	\label{eq:SelectionXdp}
	&X_{0n}^x(d^+) = X_0 \int_0^\infty \phi_{+1}^{(n)} \frac{\eta_\kappa-1}{i\varepsilon_\kappa} \kappa \dd{\kappa} \, , \\ 
	\label{eq:SelectionXdn}
	&X_{0n}^x(d^-) = 0 \, . 
\end{eqnarray} \endnumparts 
Here, $X_0=mv_F\pi/(2a^2)$ and the Fermi velocity is given by $v_F\equiv \sqrt{3}a\gamma/(2\hbar)$. The wavevector dependent functions $\eta_\kappa$ and $\varepsilon_\kappa$ are defined as $\eta_\kappa \equiv \Delta_{s\tau}/\varepsilon_\kappa$ and $2\varepsilon_\kappa \equiv \sqrt{4\Delta_{s\tau}+3\gamma^2\kappa^2}$ with $\Delta_{s\tau} \equiv \Delta+3\sqrt{3}\lambda s\tau$. 
Moreover, if the initial excitonic state possesses an angular momentum of $l$, $Q_{nm}^x$ reads 
\numparts \begin{eqnarray}
	\label{eq:SelectionQ1}
	&\frac{Q_{nm}^x(l\rightarrow l\pm 1)}{\pi a} = i \int_0^\infty \phi_{l}^{(n)} \pdv{\phi_{l\pm 1}^{(m)}}{\kappa} \kappa \dd{\kappa} \nonumber \\
	&\qquad \quad \pm i \int_0^\infty \phi_{l}^{(n)} \phi_{l\pm 1}^{(m)} (\eta_\kappa+l \pm 1) \dd{\kappa} \, , \\
	\label{eq:SelectionQ2}
	&\frac{Q_{nm}^x(l\rightarrow l\pm 2)}{\pi a} = \frac{\zeta}{2} \int_0^\infty \phi_{l}^{(n)} \phi_{l\pm 2}^{(m)}  (5\eta_\kappa-\eta_\kappa^3) \kappa \dd{\kappa} \, , \\
	\label{eq:SelectionQ3}
	&Q_{nm}^x(l\rightarrow l\pm 3) = 0 \, , \\
	\label{eq:SelectionQ4}
	&\frac{Q_{nm}^x(l\rightarrow l\pm 4)}{\pi a} = \frac{\zeta}{2} \int_0^\infty \phi_{l}^{(n)} \phi_{l\pm 4}^{(m)}  (\eta_\kappa^3-3\eta_\kappa) \kappa \dd{\kappa} \, . 
\end{eqnarray} \endnumparts 
Similar expressions can be obtained for the $y$-direction, which differ from the above equations by factors of $\pm i$. In addition, excitonic matrix elements at the K$'$ valley are found simply by interchanging plus and minus orbitals. 
These expressions clearly exhibit the excitonic selection rules as discussed in the next section.

\section{Optical selection rules \label{sec:Selection}}
Incorporation of Eqs.~(\ref{eq:SelectionXs})-(\ref{eq:SelectionXdn}) and (\ref{eq:SelectionQ1})-(\ref{eq:SelectionQ4}) into the conductivity tensors of Eqs.~(\ref{eq:Conductivity1st})-(\ref{eq:Conductivity3rd}) can be performed by utilizing the schematic diagram presented in Fig.~\ref{fig:Schematic}(a). In this diagram, the black dots represent the ground state $\st{0}$ and first few excited states $\st{s}$, $\st{p^\pm}$ and $\st{d^\pm}$.
The lines represent transitions that are allowed by the optical selection rules, whereas the missing links indicate the forbidden transitions, for which \textit{excitonic matrix elements vanish}. This diagram is obtained for the K valley, but a similar one can readily be provided for the K$'$ valley by interchanging the plus and minus states. From a physical point of view, allowed and forbidden transitions are identified by examining the conservation of the angular momentum \cite{Xiao2015}, \ie preserving the total angular momentum including the photon, lattice and exciton momenta during a light-matter interaction. If the initial and final states carry angular momenta of $l_i$ and $l_f$, respectively, the transition between them is allowed in the Dirac approximation if $l_f-l_i=\pm1$ (solid lines), while it is allowed by the TW if $l_f-l_i\pm3=\pm1$ (dashed lines). For the TW-enabled transitions, any combination of plus and minus signs is allowed, leading to a total of four possibilities. With our phase convention for Bloch eigenstates, the angular momentum of the ground state should be set to 0. Note that a $\pm$3 unit of angular momentum can be generated by the honeycomb lattice due to the $C_3$ symmetry. Moreover, a linearly-polarized photon can transfer $\pm$1 unit of angular momentum, since it is a linear combination of left- and right-handed circularly-polarized light. Using the momentum conservation rule, the transitions between $\st{0}$ and $\st{p^+}$ states are forbidden at the K valley (allowed by the TW at the K$'$ valley), whereas the $\st{s} \leftrightarrow \st{p^+}$ transitions are allowed in both valleys. Among all allowed transitions, the TW-enabled transitions are typically weaker than the rest, because they are scaled by $\zeta$. In addition, since $\eta_\kappa \approx 1$ for small $\kappa$, the matrix elements of $\st{0}\leftrightarrow\st{d^+}$ transitions are considerably smaller than $\st{0}\leftrightarrow\st{s}$ transitions, \ie $|X_{0n}^x(d^+)| < |X_{0n}^x(s)|$. Therefore, the coupling between $\st{0}$ and $\st{s}$ states is generally stronger than the remaining transitions, which is illustrated by a thicker line for $\st{0} \leftrightarrow \st{s}$ transitions in Fig.~\ref{fig:Schematic}.


Using the diagram in Fig.~\ref{fig:Schematic}(a), we can systematically identify the dominant transitions for linear and NLO processes. The many-body system is initially in its ground state with all electrons occupying the valence band. Depending on the number of photons $N$, involved in the NLO process, precisely $N$ intermediate states are visited before the system eventually return to the ground state 
Therefore, two, three and four intermediate states are involved for the first-, second-, and third-order processes, respectively, with the successive transitions given by: First order $\st{0} \rightarrow \st{m} \rightarrow \st{0}$, Second order $\st{0} \rightarrow \st{m} \rightarrow \st{n} \rightarrow \st{0}$, Third order $\st{0} \rightarrow \st{m} \rightarrow \st{n} \rightarrow \st{l} \rightarrow \st{0}$. In each case, $\st{l}$, $\st{n}$ and $\st{m}$ can be the ground or any of the excited states. Hereafter, any set of possible transitions is referred to as a (transition) \textit{path}.

\begin{figure}[t]
	\centering
	\includegraphics[width=0.45\textwidth]{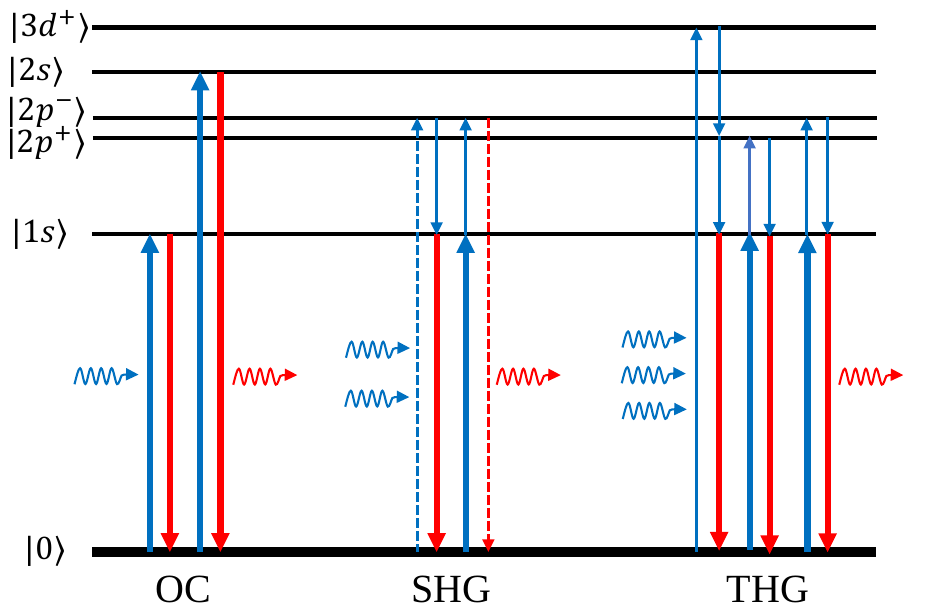}
	\caption[Exciton states]{Sketch of exciton energy levels at the K valley with the dominant transition paths (solid/dashed arrows) for the first resonances of OC, SHG and THG spectra. Dashed arrows represent the TW-enabled transitions, see Fig.~\ref{fig:Schematic}. Red arrows indicate transitions that relaxes the system to its ground state.  }
	\label{fig:ExcitonStates}
\end{figure}

For a first-order process (such as linear absorption), the dominant transition path is $\st{0} \rightarrow \st{s} \rightarrow \st{0}$, and hence, mainly $s$ excitons are bight in the OC spectrum \cite{Zhang2018}. In addition to $s$ excitons, $d^+$ and $p^-$ excitons can also be excited, but their brightness is much lower than the $s$ ones, see Eqs.~(\ref{eq:SelectionXs})-(\ref{eq:SelectionXdn}). 
Figure~\ref{fig:Schematic}(c) shows the dominant transition paths for the second-order processes (such as SHG and optical rectification), \ie S1: $\st{0} \rightarrow  \st{p^-} \rightarrow \st{s} \rightarrow \st{0}$ and S2: $\st{0} \rightarrow \st{s} \rightarrow \st{p^-} \rightarrow \st{0}$. Note that the S1 and S2 paths lead to resonances at different frequencies, since the system is relaxed through distinct states, \ie $\st{s}$ and $\st{p^-}$ for S1 and S2, respectively. The quadratic process requires at least one transition, which is enabled by the TW and hence, any quadratic response in monolayer TMDs vanishes if the TW is not included \cite{Taghizadeh2019}. Besides these two dominant paths, two other paths are conceivable for the second-order process, $\st{0} \rightarrow  \st{d^+} \rightarrow \st{s} \rightarrow \st{0}$ and $\st{0} \rightarrow  \st{s} \rightarrow \st{d^+} \rightarrow \st{0}$. However, their contributions are considerably smaller then the S1 and S2 paths due to weak coupling between the ground state and $d$ excitons, as well as between the $s$ and $d^+$ excitons [see Eq.~(\ref{eq:SelectionQ2})].
Moving to the third-order processes (such as THG or optical Kerr), four states are involved and, hence, the number of possible paths grows considerably. However, we can still identify the dominant paths as shown with labels in Fig.~\ref{fig:Schematic}(c). In contrast to the quadratic response, the third-order response does not necessitate the TW to survive. 
The present algorithm for identifying the dominant transitions can readily be extended to higher order NLO processes in monolayer TMDs.

Combining the above-mentioned selection rules with the information about exciton energies, we can predict the spectral position and nature of resonance peaks appearing in OC, SHG and THG spectra. Starting with the linear optical process, it is expected that the first strong resonances in the OC spectrum correspond to $1s$ and $2s$ states as shown schematically in Fig.~\ref{fig:ExcitonStates}. This becomes more complicated for the NLO processes since contributions of all intermediate transitions should be summed to obtain the full spectrum, \eg all transitions between $s$ and $p^-$ excitons should be considered for determining the SHG spectrum. Nonetheless, spectrally-neighboring states dominate, due to the energy-dependent denominators of Eq.~(\ref{eq:Conductivity2nd}) or (\ref{eq:Conductivity3rd}). Therefore, the lowest frequency resonances in the SHG spectrum are anticipated to be at frequencies coinciding with half of $1s$ and $2p^-$ energies, and they are dominated by the transitions indicated in Fig.~\ref{fig:ExcitonStates}. Note that probing $p$ excitons in the linear optical spectrum of monolayer TMD is challenging due to the weak coupling between $\st{0}$ and $\st{p^-}$, whereas it can be explored experimentally using the SHG response as already demonstrated in Ref.~\cite{Ye2014}. Proceeding to the THG, we expect that the lowest frequency resonance occurs at one-third of the $1s$ state energy. In addition, multiple transitions contribute simultaneously to the resonance as illustrated in Fig.~\ref{fig:ExcitonStates}. In the next section, we numerically confirm the validity of these predictions for monolayer MoS$_2$. 

\begin{figure*}[t]
	\centering
	\includegraphics[width=0.8\textwidth]{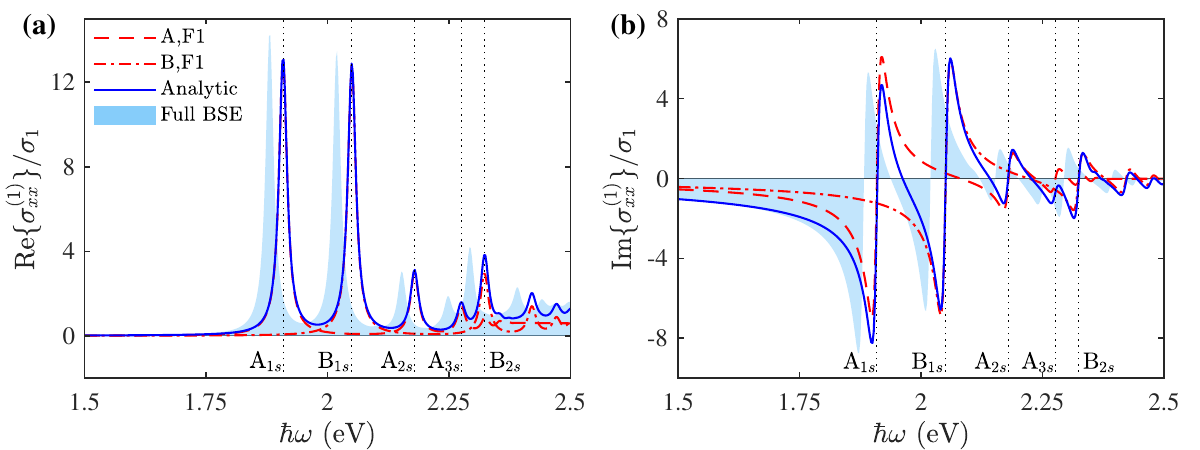}
	\caption[OC Spectrum]{Real (a) and imaginary (b) parts of excitonic OC spectrum for monolayer MoS$_2$ [$\sigma_1\equiv e^2/(4\hbar)$].  Lines show the results obtained by the analytical approach, whereas the filled area is the full BSE spectrum. Solid blue lines indicate the total response including all paths, and red dashed/dashed-dotted lines represent the contributions of A/B excitons for the F1 path [see Fig.~\ref{fig:Schematic}(b)]. }
	\label{fig:OCSpectrum}
\end{figure*}


\renewcommand{\arraystretch}{0.8}
\begin{table}[t]
	\caption{\label{tab:Excitons} Energies of the first few excitons computed using the full BSE or analytical (or Dirac) approach for monolayer MoS$_2$. }
	\begin{indented}
		\item[]\begin{tabular*}{0.99\linewidth}{@{\extracolsep{\fill} } llllll}
			\br
			Exciton & $\st{1s}$ & $\st{2p^+}$ & $\st{2p^-}$ & $\st{2s}$ & $\st{3d^+}$  \\
			\mr
			Dirac, A & 1.909 & 2.114 & 2.128 & 2.179 & 2.220 \\
			BSE, A & 1.872 & 2.088 & 2.103 & 2.151 & 2.197 \\
			\mr
			Dirac, B & 2.050 & 2.258 & 2.271 & 2.323 & 2.364 \\
			BSE, B & 2.017 & 2.228 & 2.242 & 2.291 & 2.345 \\
			\br
		\end{tabular*}
	\end{indented}
\end{table}

\section{Numerical results for monolayer MoS$_2$ \label{sec:Numerics}}
In this section, we apply the proposed approach to compute OC, SHG, and THG spectra of suspended monolayer MoS$_2$ as a representative member of the TMD family. Nonetheless, the main findings hold true for other monolayer TMDs due to similar physics. The required parameters, $\Delta$, $\gamma$ and $\lambda$, are determined by fitting to the experimental data or calculated quasi-particle band structures and, here, are set to 1.25 eV, 1.51 eV, 7.2 meV, respectively \cite{Taghizadeh2019}. A lattice constant of $a=3.18$ $\mathrm{\AA}$ and screening length of $r_0=44.3$ $\mathrm{\AA}$ are assumed \cite{Olsen2016}. Due to the symmetry of the honeycomb
lattice, at normal incidence, it is sufficient to consider the diagonal components of the conductivity tensors, \ie $\sigma_{xx}^{(1)}$, $\sigma_{xxx}^{(1)}$, and $\sigma_{xxxx}^{(1)}$ \cite{Taghizadeh2017, Hipolito2018}. The line-shape broadening is accounted for by adding a small phenomenological imaginary part, $i\Gamma$, to the frequency, i.e. $\omega \rightarrow \omega+i\Gamma$. We set $\Gamma=10$ meV for our calculations here. For the full BSE calculations, a dense $\va{k}$-mesh with more than 17000 points is employed for getting converged excitonic energies.

\begin{figure*}[t]
	\centering
	\includegraphics[width=0.8\textwidth]{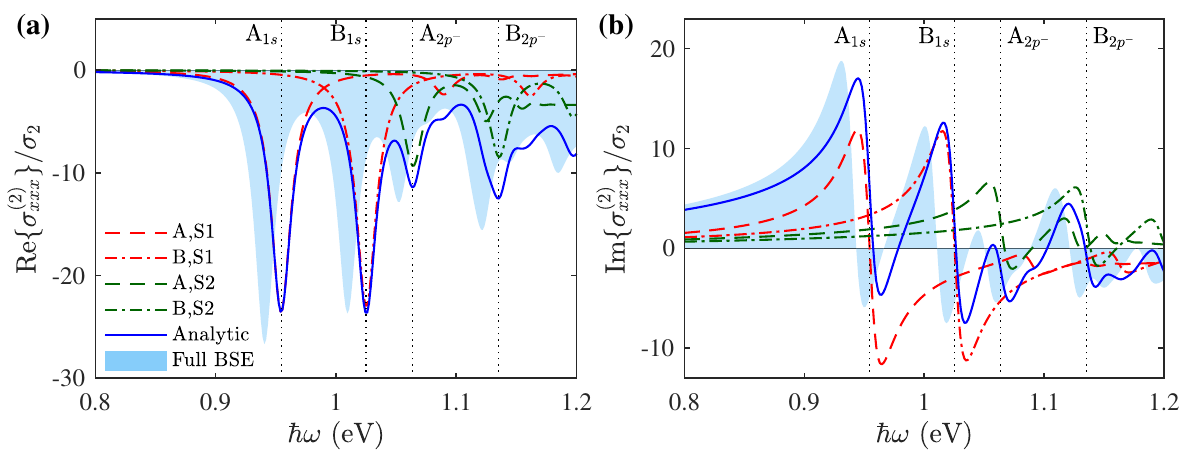}
	\caption[SHG Spectrum]{Real (a) and imaginary (b) parts of excitonic SHG spectrum for monolayer MoS$_2$ ($\sigma_2\equiv1\times 10^{-15}$ SmV$^{-1}$).  Lines show the results obtained by the analytical approach, whereas the filled area is the full BSE spectrum. Solid blue lines indicate the total response including all paths, and dashed/dashed-dotted lines represent the contributions of A/B excitons. The contributions of the S1 and S2 paths [see Fig.~\ref{fig:Schematic}(b)] are shown in red and green, respectively. }
	\label{fig:SHGSpectrum}
\end{figure*}

\begin{figure*}[t]
	\centering
	\includegraphics[width=0.8\textwidth]{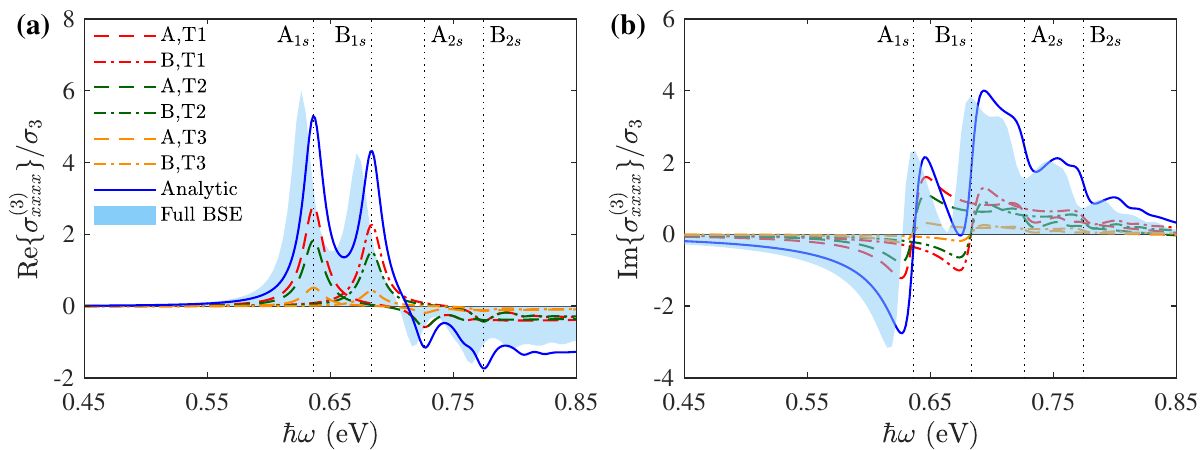}
	\caption[THG Spectrum]{Real (a) and imaginary (b) parts of excitonic THG spectrum for monolayer MoS$_2$ ($\sigma_3\equiv1\times 10^{-23}$ Sm$^2$V$^{-2}$). Lines show the results obtained by the analytical approach, whereas the filled area is the full BSE spectrum. Solid blue lines indicate the total response including all paths, and dashed/dashed-dotted lines represent the contributions of A/B excitons. The contributions of the T1, T2 and T3 paths [see Fig.~\ref{fig:Schematic}(b)] are shown in red, green, and brown, respectively.  }
	\label{fig:THGSpectrum}
\end{figure*}

We begin by comparing the exciton energies obtained using the full BSE with the ones computed by the analytical approach (in the massive Dirac approximation) in Table \ref{tab:Excitons}. In the analytical approach, Eq.~(\ref{eq:AnalyticBSE}) should be solved multiple times for different angular momenta and using two different bandgaps, $\Delta \pm 3\sqrt{3}\lambda=\{\Delta_A,\Delta_B\}$, for the A and B excitons, whereas, in the full BSE method, solving Eq.~(\ref{eq:FullBSE}) once with the TB Hamiltonian provides all required exciton energies. Nonetheless, the one-dimensional integral equation of the analytical BSE approach is solved with much less numerical effort compared to the 2D integral equation of full BSE method. Results show that the binding energy of 1s exciton is approximately $\sim$ 0.5 eV, which is in agreement with previously reported experimental \cite{Ugeda2014, Klots2014} and theoretical works \cite{Thygesen2017, Olsen2016, Huser2013, Shi2013}. Excitonic energies obtained in the analytical approach are overestimated slightly by approximately 30 meV, which is due to the underestimation of the Coulomb potential. Although this minor difference can be compensated merely by a slight reduction of the screening parameter, we use the same screening for both approaches here. Precise identification of the angular momentum characteristics of excitons obtained using the full BSE requires investigating the spatial distribution of the exciton wavefunction \cite{Ye2014}, whereas it is automatically accomplished in the analytical approach.

Upon determining the exciton energies, the OC spectrum is computed using Eq.~(\ref{eq:Conductivity1st}) and shown in Fig.~\ref{fig:OCSpectrum}. The excitonic OC spectrum displays several distinct resonances inside the bandgap due to the fundamental and higher order excitons and agrees with the previous results for monolayer MoS$_2$ \cite{Berghauser2014, Wu2015}. Apart from the minor energy shift, the analytical approach provides a very accurate replication of the BSE spectrum. Furthermore, we can readily identify the contributions of different excitons in the spectrum. The results show that the resonances in the OC spectrum mainly stem from the $s$ excitons and, hence, the first two resonances in the spectrum appear at the frequencies of the $1s$ and $2s$ states at each valley. Although transitions mediated by $p^-$ and $d^+$ excitons are allowed by the selection rules, they are barely observable in the spectrum due to their negligible contribution. Indeed, we numerically confirm that the magnitude of the matrix element between the ground state and $1s$ (or $2s$) exciton is at least 10 times larger than that of the ground state and the $2p^-$ or $3d^+$ exciton.

For the second-order processes, we limit our analysis to SHG, but similar results can be obtained for other quadratic processes such as optical rectification \cite{Taghizadeh2019}. Figure~\ref{fig:SHGSpectrum} illustrates the SHG spectrum computed for monolayer MoS$_2$ using the full BSE and analytical approaches, which shows a good agreement between the two methods. As expected for a typical excitonic SHG spectrum, excitons appear as strong resonances at frequencies below half the bandgap. The spectral positions of the fundamental peaks agree well with the experimental reports \cite{Malard2013, Trolle2015}. Using the analytical approach, we identify and label the resonance peaks according to the responsible exciton. The contributions of the two dominant paths in Fig.~\ref{fig:Schematic}(c) are distinguished. These two paths generate resonances with similar strength but at different frequencies, corresponding to the $s$ and $p^-$ exciton energies.
We also numerically examine the magnitude of the matrix element between the $1s$ and $2p^-$ or $3d^+$ states, and confirm that the former is approximately 10 times larger the latter. Combining this fact with the weak coupling of the ground state and $d^+$ excitons leads to a negligible contribution of the transition paths via the $d^+$ excitons in the total SHG response. Note that, for circularly polarized light, the resonances corresponding to $s$ and $p$ excitons possess different valley dichroism due to their distinct angular momenta \cite{Taghizadeh2019}.


In Fig.~\ref{fig:THGSpectrum}, we compare the THG conductivity computed using the full BSE with the analytical approach. The THG results follow the main trends observed in the OC and SHG responses. However, multiple transition paths should be included in order to obtain an accurate spectrum. In contrast to the OC and SHG responses, where each resonance is mainly formed by a single transition path, the resonances in THG spectrum have contributions from several distinct paths. For instance, the fundamental peaks, \ie A$_{1s}$ and B$_{1s}$, originate from three paths; T1: $\st{0} \rightarrow \st{s} \rightarrow \st{p^-} \rightarrow \st{s} \rightarrow \st{0}$, T2: $\st{0} \rightarrow \st{s} \rightarrow \st{p^+} \rightarrow \st{s} \rightarrow \st{0}$, and T3: $\st{0} \rightarrow \st{d^+} \rightarrow \st{p^+} \rightarrow \st{s} \rightarrow \st{0}$. Note that, although the T3 path includes a weak transition between the ground and $d^+$ states, the coupling between $d^+$ and $p^+$ states can be sufficiently large so as to compensate this weak transition and, hence, contribute significantly to the THG response. Indeed, our calculations show that the magnitude of the coupling between the $2p^+$ and $3d^+$ states is almost 3 times larger than the coupling between the $1s$ and $2p^+$ states. 



\section{Conclusion \label{sec:Conclusion}}
In summary, we have elucidated an analytical approach for calculating the linear and NLO responses of intrinsic monolayer TMDs, including excitonic effects. Analytical expressions for excitonic matrix elements were derived and used to determine the NLO response of monolayer MoS$_2$. With a much less numerical effort, the analytical approach generates almost identical spectra to that of the numerically demanding BSE, while it also provides a transparent picture of the excitonic transitions in the NLO response. Our approach can readily be extended to identify the main transitions for any other NLO process in monolayer TMDs and, hence, provides a simple but accurate tool for analyzing experimental NLO spectra of monolayer TMDs.

\ack
The authors thank F. Hipolito for helpful discussions throughout the project. 
This work was supported by the QUSCOPE center sponsored by the Villum Foundation and the CNG center under the Danish National Research Foundation (project DNRF103).

\appendix

\section{Equation of motion and its perturbative solution \label{sec:AppendixA}}
Here, we review the derivation of the dynamical equation for the density matrix and present its perturbative solution up to third order in the external field. We follow the procedure outlined in Refs.~\cite{Pedersen2015, Taghizadeh2018}. For a two-band semiconductor, the dynamical equations for the population densities and coherences in the mean-field and Tamm-Dancoff approximations read \cite{Pedersen2015, Taghizadeh2018}
\numparts \begin{eqnarray} 
	\label{eq:DensityDiffEqA}
	&\fl i\hbar\pdv{\rho_{cv\va{k}}}{t} - \varepsilon_{cv\va{k}} \rho_{cv\va{k}} + \sum_{\va{k}'} V_{cv}^{\va{k},\va{k}'} (\rho_{vv\va{k}}-\rho_{cc\va{k}}) \rho_{cv\va{k}'} \nonumber \\
	& \qquad -\sum_{\va{k}'} \big[ V_{vv}^{\va{k},\va{k}'} (\rho_{vv\va{k}'}-1) - V_{cc}^{\va{k},\va{k}'} \rho_{cc\va{k}'} \big] \rho_{cv\va{k}} \nonumber \\
	& \qquad =(u_{cc\va{k}}-u_{vv\va{k}}) \rho_{cv\va{k}} + u_{cv\va{k}} (\rho_{vv\va{k}}-\rho_{cc\va{k}}) \, , \\
	\label{eq:DensityDiffEqB}
	&\fl i\hbar\pdv{\rho_{cc\va{k}}}{t} - \sum_{\va{k}'} \big[ V_{vc}^{\va{k},\va{k}'} \rho_{vc\va{k}'} \rho_{cv\va{k}} - V_{cv}^{\va{k},\va{k}'} \rho_{cv\va{k}'} \rho_{vc\va{k}} \big] \nonumber \\
	& \qquad \qquad \qquad =u_{cv\va{k}}\rho_{vc\va{k}}-u_{vc\va{k}}\rho_{cv\va{k}} \, . 
\end{eqnarray} \endnumparts
Here, $u_{nm\va{k}} \equiv \langle n\va{k}| \hat{u}(t) |m\va{k} \rangle$ is the single-particle matrix element of the light-matter interaction Hamiltonian $\hat{u}(t)$. In the dipole approximation (long wavelength regime), the single-electron interaction Hamiltonian reads $\hat{u}(t) \equiv e\hat{\va{r}} \cdot \bmcal{E}(t)$. In Eqs.~(\ref{eq:DensityDiffEqA}) and (\ref{eq:DensityDiffEqA}), we ignore the exchange terms due to their negligible influence \cite{Thygesen2017}. Moreover, $V_{nm}^{\va{k},\va{k}'}$ are the direct Coulomb matrix elements, given by  
\begin{equation}
\label{eq:CoulombMatrix}
	V_{nm}^{\va{k},\va{k}'} \approx \frac{e^2}{2\epsilon_0} \frac{\langle n\va{k}|n\va{k}'\rangle \langle m\va{k}'|m\va{k}\rangle }{\abs{\va{k}-\va{k}'}(\epsilon_s+r_0\abs{\va{k}-\va{k}'})} \, , 
\end{equation}
where a Keldysh potential with substrate screening of $\epsilon_s$ and screening length of $r_0$ is assumed \cite{Taghizadeh2019}. The approximation in Eq.~(\ref{eq:CoulombMatrix}) is mainly because of the neglected terms for non-zero reciprocal vectors \cite{Taghizadeh2018}. 
The equation of motion for $\rho_{cv\va{k}}$ is simply obtained by complex conjugating the first equation, since the density matrix operator is Hermitian. In addition, the population conservation, \ie normalization of the density matrix, leads to $\rho_{vv\va{k}} = 1-\rho_{cc\va{k}}$.

For a time-harmonic electric field, \ie $\hat{u}(t)=\hat{u}_\omega \exp(-i\omega t) + \mathrm{h.c.}$, the dynamical equation for the density matrix can be solved perturbatively up to any required order of the field using the Green's functions \cite{Pedersen2015}.
Here, we only focus on the expressions for second/third harmonic generation for brevity, but similar results can be readily obtained for the optical rectification and Kerr processes. Hence, the $N^{th}$-order density matrix reads $\rho_{ij\va{k}}^{(N)}(t) = \rho_{ij\va{k}}^{(N\omega)} \exp(-iN\omega t) + \rho_{ij\va{k}}^{(-N\omega)} \exp(iN\omega t)$, where $\rho_{ij\va{k}}^{(N\omega)}$ for $N=1,2,3$ are given by
\numparts \begin{eqnarray}
	\label{eq:DensityFinalA}
	\rho_{cv\va{k}}^{(\omega)}  = \sum_n \frac{\psi_{\va{k}}^{(n)} U_{n0}}{\hbar \omega-E_n} \, , \\
	\rho_{cv\va{k}}^{(2\omega)} = \sum_{n,m} \frac{\psi_{\va{k}}^{(n)} U_{nm} U_{m0}}{(2\hbar\omega-E_n)(\hbar\omega-E_m)} \, , \\
	\rho_{cv\va{k}}^{(3\omega)} = \sum_{l,n,m} \frac{\psi_{\va{k}}^{(l)} U_{ln} U_{nm} U_{m0}}{(3\hbar\omega-E_l)(2\hbar\omega-E_n)(\hbar\omega-E_m)} \, , \\
	\rho_{cc\va{k}}^{(\omega)} = 0 \, , \\
	\rho_{cc\va{k}}^{(2\omega)} = \rho_{cv\va{k}}^{(\omega)} \rho_{vc\va{k}}^{(\omega)} \, , \\
	\label{eq:DensityFinalF}
	\rho_{cc\va{k}}^{(3\omega)} = \rho_{cv\va{k}}^{(2\omega)} \rho_{vc\va{k}}^{(\omega)} + \rho_{cv\va{k}}^{(\omega)} \rho_{vc\va{k}}^{(2\omega)} \, .
\end{eqnarray} \endnumparts
Here, $U_{0n} = U_{n0}^* \equiv \sum_{\va{k}} \psi_{\va{k}}^{(n)} u_{vc\va{k}}^{(\omega)}$ and $U_{nm} \equiv \sum_{\va{k}} \psi_{\va{k}}^{(n)*} \psi_{\va{k}}^{(m)} [u_{cc\va{k}}^{(\omega)} - u_{vv\va{k}}^{(\omega)}]$ are the excitonic matrix elements of the interaction Hamiltonian, where $u_{nm\va{k}}^{(\omega)} \equiv \ev{n\va{k}}{\hat{u}_\omega}{m\va{k}}$. Indeed, $U_{0n}$ corresponds to the coupling between the ground state $\st{0}$ and excited state $\st{n}$, whereas $U_{nm}$ is the inter-excitonic matrix element.
Note that $\rho_{vc\va{k}}^{(N\omega)}=\Big[\rho_{cv\va{k}}^{(-N\omega)}\Big]^*$ and $\rho_{vv\va{k}}^{(N\omega)} = -\rho_{cc\va{k}}^{(N\omega)}$ due to the Hermiticity and normalization of the density matrix, respectively. Upon obtaining the density matrix, the expectation value of the $N^{th}$-order current density is determined straightforwardly as $\va{J}^{(N)}(t)=\mathrm{Tr}[\hat{\rho}^{(N)}\hat{\va{J}}]$ leading to Eqs.~(\ref{eq:Conductivity1st})-(\ref{eq:Conductivity3rd}) for the conductivities. 

\section{Trigonal-warping Hamiltonian \label{sec:AppendixB}}
For the 2-by-2 TB Hamiltonian, Eq.~(\ref{eq:TBHamiltonian}), the eigenenergies for a given spin read $\varepsilon_{c\va{k}} = -\varepsilon_{v\va{k}} = \sqrt{\Delta_s^2+\gamma^2 F^2}$ with $\Delta_s \equiv \Delta+s \lambda g(\va{k})$. The corresponding eigenvectors are then given by $|c\va{k} \rangle = [\cos(b/2), -\sin(b/2)e^{i\phi}]$ and $|v\va{k} \rangle = [\sin(b/2), \cos(b/2)e^{i\phi}]$, where $F$ and $\phi$ are defined using $f(\va{k})=F \exp(i\phi)$, and $\cos(b) \equiv \Delta_s/\varepsilon_{c\va{k}}$. Furthermore, the momentum operator in the TB method is given as the $\va{k}$-derivative of the Hamiltonian, $\hbar p^\alpha=m \pdvi{H_0 }{k_\alpha}$. Hence, the momentum matrix elements can readily be determined using $p_{nm\va{k}}^\alpha=\ev{n\va{k}}{p^\alpha}{m\va{k}}$, \eg $\hbar p_{vc\va{k}}^\alpha= -m\gamma[\cos(b)F_\alpha+iF\phi_\alpha]$ with $F_\alpha\equiv \pdvi{F}{k_\alpha}$ and $\phi_\alpha\equiv \pdvi{\phi}{k_\alpha}$.

Taylor-expanding $f(\va{k})$ and $g(\va{k})$ functions in the vicinity of the Dirac points, K: $2\pi[3^{-1/2},3^{-1}]/a$ and K$'$: $2\pi[3^{-1/2},-3^{-1}]/a$, the massive Dirac and TW Hamiltonian are determined. Using Eq.(\ref{eq:TWFunction}) for $f(\va{k})$, we calculate any required parameter and keep only 0$^\mathrm{th}$ and 1$^\mathrm{st}$ order terms in $\zeta$. Hence, the transition energies, $\varepsilon_{cv\va{k}}$, momentum matrix elements, $p_{vc\va{k}}^\alpha$, and Berry connections, $\Omega_{cc\va{k}}^\alpha-\Omega_{vv\va{k}}^\alpha$, are given by  
\numparts \begin{eqnarray}
	\label{eq:TransitionEnergy}
	&\varepsilon_{cv\va{k}} \approx 2\varepsilon_\kappa\big[1- \zeta \tau \kappa (1-\eta_\kappa^2) \sin(3\theta) \big] \, , \\
	\label{eq:xMomentumMatrix}
	&\frac{p_{vc\va{k}}^x}{p_0} \approx \eta_\kappa\cos(\theta) + i\tau\sin(\theta) + \frac{\zeta\kappa}{2} \bigg[ 3i\cos(2\theta) -i\cos(4\theta) \nonumber \\
	& \quad -\tau\eta_\kappa(4+\eta_\kappa^2)\sin(2\theta) + \tau\eta_\kappa(2-\eta_\kappa^2)\sin(4\theta) \bigg] \, , \\
	\label{eq:yMomentumMatrix}
	&\frac{p_{vc\va{k}}^y}{p_0} \approx \eta_\kappa\sin(\theta) - i\tau\cos(\theta) - \frac{\zeta\kappa}{2} \bigg[ 3i\sin(2\theta) + i\sin(4\theta) \nonumber \\
	& \quad +\tau\eta_\kappa(4+\eta_\kappa^2)\cos(2\theta) +\tau\eta_\kappa(2-\eta_\kappa^2)\cos(4\theta) \bigg] \, , \\
	\label{eq:xBerry}
	&\frac{\Omega_{cc\va{k}}^x-\Omega_{vv\va{k}}^x}{a} \approx \eta_\kappa \bigg[ \tau\frac{\sin(\theta)}{\kappa}+  \frac{\zeta}{2}(5-\eta_\kappa^2)\cos(2\theta) \nonumber \\
	&\qquad \qquad \qquad -\frac{\zeta}{2}(3-\eta_\kappa^2)\cos(4\theta)\bigg] \, , \\
	\label{eq:yBerry}
	&\frac{\Omega_{cc\va{k}}^y-\Omega_{vv\va{k}}^y}{a} \approx -\eta_\kappa \bigg[\tau\frac{\cos(\theta)}{\kappa}+\frac{\zeta}{2}(5-\eta_\kappa^2)\sin(2\theta) \nonumber \\
	&\qquad \qquad \qquad +\frac{\zeta}{2}(3-\eta_\kappa^2)\sin(4\theta)\bigg] \, ,
\end{eqnarray} \endnumparts
with $p_0\equiv-mv_F$ . 
In these expressions, the terms proportional to $\zeta$ stem from the TW effect, and well-known expressions in the massive Dirac approximation are obtained by setting $\zeta=0$.  

\section{Analytical BSE \label{sec:AppendixC}}
In the massive Dirac approximation, a simple expression can be obtained for the Coulomb matrix element.
Then, using an ansatz $\psi_{\va{k}}^{(n)} = \exp(il\theta ) \phi_{l}^{(n)}(\kappa)$ and performing the angular integral, the BSE is reduced to the following 1D eigenvalue problem for the radial part of exciton wavefunction \cite{Wu2015}:  
\begin{eqnarray}
	\label{eq:AnalyticBSE}
	&E_n \phi_{l}^{(n)}(\kappa) = 2\varepsilon_\kappa \phi_{l}^{(n)}(\kappa) - \frac{e^2}{8(2\pi)^2\epsilon_0 a} \int_{0}^\infty \dd{\kappa'} \nonumber \\
	& \qquad \times \phi_l^{(n)}(\kappa') \kappa'  \Big\{2\sin(b_\kappa) \sin(b_{\kappa'})D_l(\kappa,\kappa') \nonumber \\ 
	& \qquad +[1+\cos(b_\kappa)][1+\cos(b_{\kappa'})]D_{l+\tau}(\kappa,\kappa') \nonumber \\
	& \qquad + [1-\cos(b_\kappa)][1-\cos(b_{\kappa'})]D_{l-\tau}(\kappa,\kappa') \Big\} \, . 
\end{eqnarray}
Here, $\cos(b_\kappa)\equiv\Delta_{s\tau}/\varepsilon_\kappa$ and $D_l(\kappa,\kappa')$ is defined as
\begin{equation}
	D_l(\kappa,\kappa') \equiv \int_0^{2\pi}  \frac{\cos(l\theta)\dd{\theta}}{q(\epsilon_s+r_0q/a)} \, ,
\end{equation}
with $q\equiv \sqrt{\kappa^2+\kappa'^2-2\kappa\kappa'\cos(\theta)}$. From Eq.~(\ref{eq:AnalyticBSE}), one can confirm that the $l$ state at one valley is degenerate with the $-l$ state at the other valley [note that $D_l(\kappa,\kappa')=D_{-l}(\kappa,\kappa')$]. For each angular momentum, Eq.~(\ref{eq:AnalyticBSE}) can be solved numerically by discretizing $\kappa$, using the technique explained in Ref.~\cite{Chao1991}. Upon determining the exciton eigenenergies and eigenvectors, the excitonic matrix elements are obtained using Eqs.~(\ref{eq:SelectionXs})-(\ref{eq:SelectionXdn}) and (\ref{eq:SelectionQ1})-(\ref{eq:SelectionQ4}).


\section*{References}
\bibliographystyle{unsrt}
\bibliography{Selection}

\end{document}